\documentclass[showpacs,twocolumn,prb,eqsecnumbers]{revtex4}
\usepackage{amsmath}
\usepackage{epsf}
\usepackage{graphicx}
\usepackage{dcolumn}
\usepackage{bm}

\newcommand{\nn}{\nonumber}
\newcommand{\beq}{\begin{equation}}
\newcommand{\eeq}{\end{equation}}
\newcommand{\bea}{\begin{eqnarray}}
\newcommand{\eea}{\end{eqnarray}}
\newcommand{\bwt}{\begin{widetext}}
\newcommand{\ewt}{\end{widetext}}

\newcommand{\bk}{{\mathbf k}}
\newcommand{\bp}{{\mathbf p}}
\newcommand{\bq}{{\mathbf q}}

\begin{document}

\title{Spin conservation and Fermi liquid
 near a ferromagnetic quantum critical point}
\author{Andrey V. Chubukov$^{1,3}$ and Dmitrii L. Maslov$^{2,3}$}
\date{\today}

\begin{abstract}
We propose a new low-energy theory for itinerant fermions near 
a  ferromagnetic quantum critical point.
We show that the full low-energy model includes, in addition to conventional interaction via spin fluctuations, 
another type of interaction, whose presence is crucial for the theory to satisfy
 $SU(2)$ spin conservation. We demonstrate the consistency between a loop-wise expansion
 and a Fermi liquid description for the full model.  We further show that,
 prior to the ferromagnetic instability, the system develops a Pomeranchuk-type instability into a state with zero magnetization but with $p$-wave deformations of the Fermi surfaces of spin-up and -down electrons 
(a 
spin nematic).

\end{abstract}
\affiliation{
$^{1}$
 Department of Physics, University of Wisconsin-Madison, 1150 University
Ave., Madison, WI 53706-1390\cite{perm}
\\
$^{2}$ Department of
Physics, University of Florida, P. O. Box 118440, Gainesville, FL
32611-8440\cite{perm}
\\
$^{3}$Max-Planck-Institut f\"{u}r Physik komplexer Systeme, D-01187
 Dresden, Germany}
\maketitle

{\it Introduction}~~~~
Conservation laws are powerful tools to study  the behavior of
interacting fermions. The Hamiltonian 
 for a system of fermions with non-relativistic  interactions is $SU(2)$ invariant and preserves the
total charge and spin of particles. 
 In this Letter,
 we analyze the consequences
 of spin conservation 
 for  effective low-energy theories, which
   describe
 the behavior of itinerant systems
 near a  $q=0$ Pomeranchuk 
instability.
 In particular, we focus on  
  the effective theory of a
 ferromagnetic quantum critical point
 (FM QCP).
It has been long
assumed
 that a FM QCP
  is well described by the spin-fermion model ~\cite{hertz} (SFM) with interaction only in the spin channel,
 mediated by small $q$ 
   collective spin excitations
 (for a review, see 
 Refs.~\onlinecite{scal,rosch_rmp}). 
 
Here, we report two results. First, we show that
 the SFM is fundamentally incomplete as it
 violates 
 spin conservation. 
  We obtain 
 a proper low-energy model which obeys
 spin conservation and show that in this new model the 
interaction is mediated by {\it both} spin and charge bosons;
 the charge component of the interaction 
is of the same strength as the  spin component.
 We show that the perturbation theory for this model  is consistent with 
 the Fermi liquid (FL) description, which is based on conservation laws.
 This consistency does not hold in the SFM, as it has been recently emphasized in Ref.~\onlinecite{khodel}.

Second, we show that, within the new model, the system develops an 
instability toward a $p-$wave spin nematic state,\cite{hirsch,wu} in which the net magnetization is absent but the Fermi surfaces of spin-up and spin-down electrons are shifted in opposite directions. For 
not too long-range 
 interaction 
in the spin channel, this instability occurs not only before the Stoner instability into a ferromagnetic state but also before other known instabilities near a FM QCP --a first order transition, helical magnetic order, or
  p-wave superconductivity. ~\cite{belitz_rmp,rosch_rmp,chi_last,p-wave,green}

A $p-$wave spin nematic state,
first introduced by 
Hirsch,~\cite{hirsch}
 has been   
 studied in detail in recent years.~\cite{wu}
  It 
  was identified as 
  the ground state
  of a  particular lattice model,~\cite{hirsch} 
  but  whether such a state 
  can be realized in a rotationally
 and $SU(2)$-spin invariant model was not clear. Our result shows that
such a state emerges in an ordinary isotropic FL tuned to a FM QCP.

\begin{figure}[tbp]
\caption{a) Ladder diagrams for $\Gamma^\Omega$. Summing up these diagrams for a contact interaction $U$ (dashed line) and
 re-arranging spin indices, one obtains 
 the effective spin-spin interaction near the critical point (wavy line), where $Um/2\pi\approx 1$. 
 The rest of the diagrams (not shown) eliminate the instability in the charge channel. 
b) 
Renormalization of the spin-fermion vertex.
The last two diagrams are the Aslamazov-Larkin--type (AL) contributions 
which restore spin conservation.
 c) The full effective interaction $\Gamma^{\Omega,\mathrm{full}}$ with contributions
 from the direct spin-spin interaction and AL-type   processes. }
 \label{fig1}
\begin{center}
\epsfxsize=1.0\columnwidth
\epsffile{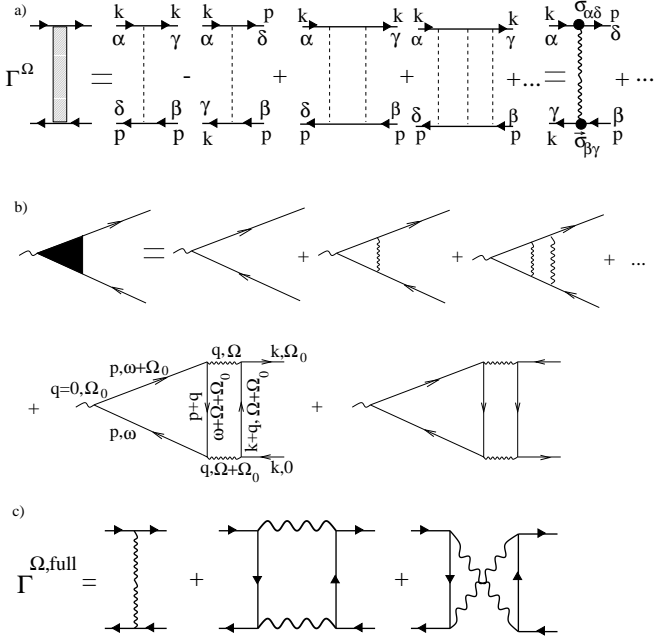}
\end{center}
\end{figure}

{\it Spin conservation in the SFM.}~~~~
The SFM model was introduced as a minimal model near a magnetic
 QCP.~\cite{hertz,acs}
It describes  low-energy fermions with an effective
spin-spin interaction 
\beq
\Gamma^\Omega_{\alpha\beta;\gamma\delta} (
\bk,\omega_k;\bp,\omega_p)
 = \frac{1}{2\nu Z} 
 \frac{1}{\delta + (a q)^2 + \frac{|\Omega|}{v_F q}}
{\vec \sigma}_{\alpha\delta}  \cdot{\vec \sigma}_{\beta\gamma}.
\label{2_1}
\eeq 
Here, 
${\bf q} = {\bf k} - {\bf p}, ~\Omega = \omega_{k} - \omega_{p}$ 
 are the relative momenta and Matsubara frequencies, respectively,
 $\delta$ measures the distance from the QCP, $\nu$
 is the   density of states, and $Z$ is 
the fermionic residue.
  The 
 superscript $\Omega$ implies that Eq.~(\ref{2_1}) is obtained  by 
 integrating out fermions outside the Fermi surface (FS) but contains 
 no contribution  from fermions at the FS -- the latter is obtained within 
  the FL theory which takes $\Gamma^\Omega$  an input.~\cite{agd}
 The main structure of Eq. (\ref{2_1}) is obtained already within the 
 RPA~\cite{scal,we,berk} (see Fig. \ref{fig1}a); 
 an additional factor of $1/Z$ arises from renormalization of $\Gamma^{\Omega}$ beyond the RPA. \cite{we}

The effects of the interaction $\Gamma^\Omega$ can be analyzed in two ways --
 either within a loop-wise expansion in
$\Gamma^\Omega$ 
or with the FL theory built upon conservation laws.  
In a model which satisfies conservation laws, the results of the two approaches must be identical. 
We first show that they are not in the SFM. 
 For definiteness,  we  consider below the case of $D=2$,
 but our results are also  valid for $D=3$. Loop-wise expansions 
  of bosonic and fermionic self-energies
  have 
 been performed by a number of authors,
  and the results  relevant to our discussion are~\cite{aim,chub_cross} 
\beq
\frac{m^*}{m} \approx \frac{1}{Z} = \lambda \equiv \frac{3}{4ak_F \sqrt{\delta}}; ~
\chi_s = \chi^0_s \propto \frac{1}{\delta}.
\label{n_1}
\eeq
Observe that the static uniform spin susceptibility is not renormalized by the interaction at low energies.
 
To obtain the results for $m^*/m$ and $\chi_s$ in 
 the FL theory, one 
 has to convert $\Gamma^\Omega$ into 
  the
 Landau function $g_{\alpha\beta;\gamma\delta}
 (\bk,\bp) 
= Z^2 (m^*/m) \Gamma^\Omega_{\alpha\beta;\gamma\delta} 
 (\bk,0;\bp,0)
$, 
 obtain
 charge and spin harmonics of 
$g_{\alpha\beta;\gamma\delta} = g_c \delta_{\alpha \gamma} \delta_{\beta \delta} + g_s {\vec\sigma}_{\alpha \gamma}\cdot {\vec \sigma}_{\beta \delta}$  for the particles on the FS ($|{\bf k}| = |{\bf p}| = k_F,~\omega_k = \omega_p =0$), 
 and use 
  familiar FL relations 
 $m^*/m = 1 + g_{c,1}$ and $\chi_s = \chi^0_s (1 + g_{c,1})/(1 + g_{s,0})$. 
 Carrying out this procedure, we find $g_{c,1} = \lambda$ and $g_{s,0} = -\lambda/3$, i.e.,
\beq
\frac{m^*}{m} = \lambda; ~\chi_s  = \chi^0_s \frac{1 + \lambda}{1 - \frac{\lambda}{3}}.
\label{n_2}
\eeq
 The result for $m^*/m$ is consistent with Eq.~(\ref{n_1}) but the result for $\chi_s$ is not.
  Moreover, Eq. ~(\ref{n_2}) shows that $\chi_s$ 
  has a spurious divergence at 
  $\lambda =3$, i.e., at finite distance from the QCP. 

This disagreement between the two approaches is related to a violation of spin conservation in the SFM. 
 To see this,  we compute 
 the
 spin and charge susceptibilities in the SFM at a small but finite frequency $\Omega$ and
   zero momentum $q$. The
 Ward identities require that $\chi_{c,s} (q=0, \Omega)=0$.
We
  show that 
  $\chi_{s} (q=0, \Omega)\neq 0$, although  $\chi_{c} (q=0, \Omega)=0$.

  The effect of the interaction with collective 
 modes  on the spin and charge susceptibilities
 can be parameterized as
  \beq
\chi_{c,s} (q=0, \Omega) =
 \chi^{0}_{c,s} \left(1 - \frac{\Lambda_{c,s}}{1 + \lambda}\right),
\label{1}
\eeq
where $\Lambda_{c,s}$ are the
 fully renormalized fermion-boson vertices normalized to 
$\Lambda_{c,s} =1$ for non-interacting fermions. Self-consistency requires that 
$\Lambda_{c,s}$ 
 are obtained within the same 
 RPA
 approximation as Eq.~(\ref{2_1}) itself.  The 
 RPA 
 series  [cf. Fig.~ \ref{fig1}$b$)] 
 is geometric;
  the $n$-th order term  is given by
$(\lambda/(1+\lambda))^n$ for the charge vertex and $(-1/3)^n (\lambda/(1+\lambda))^n$ for the spin vertex.~\cite{chub_ward} The difference comes from 
 spin summation:  the charge vertex contains a combination $\delta_{\alpha\beta} {\vec \sigma}_{\alpha \gamma}\cdot {\vec \sigma}_{\delta \beta} = 3 \delta_{\gamma \delta}$, while the spin vertex contains
$\sigma^i_{\alpha\beta} {\vec \sigma}_{\alpha \gamma}\cdot {\vec \sigma}_{\delta \beta} = - \sigma^i_{\gamma \delta}$. Summing up
 the series, we find that
 $\Lambda_{c} =
  1 + \lambda$ and
 $\Lambda_{s} = 
 (1 + \lambda)/(1 + 4 \lambda/3)$. Substituting 
  $\Lambda_{c,s}$ into Eq.~(\ref{1}), we find that charge susceptibility $\chi_s (q=0, \omega)$ vanishes, as it should, but $\chi_s (q=0, \omega)$ does not, 
 which implies that spin conservation is violated.

This violation has a physical  explanation.\cite{chub_ward}  
In the SFM, electron spins are split into spins of itinerant electrons, 
${\bf s} = c^{\dagger}_\alpha {\vec \sigma}_{\alpha\beta} c_\beta$ 
 and spins ${\bf S}$  of collective bosons.
 In the $SU(2)$ 
 symmetric
 case, the fermion-boson coupling
 ${\bf s} \cdot{\bf S}$  flips ${s}^z$.
 As a result, ${\bf s}$ is not conserved separately  from ${\bf S}$.

Because of spin non-conservation, the  SFM
 was argued~\cite{chub_ward} to be applicable only in the regime when conservation laws do not matter, i.e.,  when relevant momenta are much larger than relevant frequencies. 
 Some physical properties described by the SFM, e.g., mass renormalizaton, superconducting and other instabilities of a paramagnetic state, 
 come from 
fermions from that range and are therefore 
adequately described within the SFM.\cite{chub_cross,chi_last} Still, the violation of spin conservation 
 implies that the SFM model is fundamentally incomplete as the $SU(2)$ invariance is not broken
in a paramagnetic phase at any energy. A correct low-energy model,
which includes all relevant interactions,
 should obey spin conservation.  

{\it Spin-conserving low-energy model.}~~~~
We now construct such a
 model by going beyond the ladder approximation
for the vertex corrections  $\Lambda_c$ and $\Lambda_s$.
  It turns out that the important diagrams are the Aslamazov-Larkin (AL) type-diagrams encountered, e.g., in the theory of superconducting fluctuations.\cite{al} 
  To second order, there  are two 
  such diagrams 
  shown in the last line of
  Fig.~\ref{fig1}
$b)$. For the charge vertex, the two diagrams cancel each other. \cite{comm_2} For the spin vertex,
 they add up and contribute 
\bwt
\bea
I_{\mathrm{AL}}&=&\frac{1}{2\nu Z^2}\int \frac{dqq}{2\pi}\int\frac{d\Omega}{2\pi}\int \frac{d\theta_{\bp\bq}}{2\pi}\int\frac{\theta_{\bk\bq}}{2\pi}\int \frac{d\omega}{2\pi}\int d\epsilon_pG(\bp,\omega)G(\bp,\omega+\Omega_0)G(\bk_F+\bq,\Omega+\Omega_0)\nn\\&&\times \left[G(\bp+\bq,\omega+\Omega+\Omega_0)
+
 G(\bp-\bq,\omega-\Omega)\right]{\bar\Gamma}(\bq,\Omega){\bar \Gamma(\bq,\Omega+\Omega_0)}
\label{n_3}
\eea
\ewt
Here, $G(\bk,\omega)=Z/
[i\omega-\epsilon_k (m/m^*)
]$ is the Green's function of FL quasiparticles, $\epsilon_k=v_F(k-k_F)$, 
$m^*/m=Z^{-1}=1+\lambda$, 
$\theta_{\mathbf{l}\mathbf{m}}=\angle(\mathbf{l},\mathbf{m})$,
and ${\bar\Gamma}(\bq,\Omega)=\left(\delta+(aq)^2+|\Omega|/v_Fq\right)^{-1}$.
 Although $I_{\mathrm{AL}}$ is formally a two-loop correction in ${\bar\Gamma}$, its contribution is 
 actually
 of the same order as from  the one-loop correction. To see this, we observe that relevant $\epsilon_p$ in Eq.~(\ref{n_3}) are small (of order $\Omega_0$) while $q$ is finite. Hence, the integration over $\epsilon_p$ involves effectively only the first two Green's functions in the first line. For the poles of these two Green's functions to be in different half-planes, the range of $\omega$ must be bounded by the external frequency $-\Omega_0<\omega<0$. Hence, one can neglect $\omega$ and $\Omega_0$ in the remainder of the integral. The angular integrations now factorize and each of them gives a factor of 
 $(m^*/m)/v_Fq$. The remaining integrals over $q$ and $\Omega$ have a scaling property $\int dq\int d(\Omega/v_Fq) {\bar\Gamma}^2(q,\Omega/v_Fq)=\int dq{\bar\Gamma(q,0)}\sim\lambda$.  A more accurate calculation yields 
$I_{\mathrm{AL}} = (4/3) \lambda/(1 + \lambda)$ (the denominator comes from power-counting factors of $Z$).
  Combining  $I_{\mathrm{AL}}$ with the one-loop ladder vertex correction,
  which contributes $-(1/3) \lambda/(1 + \lambda)$, we find that 
 the effective one-loop correction to $\Lambda_s$ becomes $
  \lambda/(1 + \lambda)$, which is exactly the same result as the correction to $\Lambda_c$.  
 It can be readily shown that the 
 effective $n$th other contribution to $\Lambda_s$ now is 
$(\lambda/(1 + \lambda))^n$, again the same as for $\Lambda_c$. 
Summing up the series, we obtain that now $\Lambda_s = 1 + \lambda$, 
which means that $\chi_s(0,\Omega_0\to 0)=0$ and thus spin conservation is restored.
Other two-loop vertex corrections are small as
 $\max\{\Omega_{\mathrm{FL}}/E_F,\Omega_0/\Omega_{\mathrm{FL}}\}$, where $\Omega_{\mathrm{FL}}\propto \delta^{3/2}$ is the upper boundary for the FL description.

For completeness, we also analyzed spin conservation
 right at a FM QCP, where the fermionic 
   self-energy
 $\Sigma (\omega) \propto \omega^{2/3}$. The calculations at QCP are more involved, and to verify the spin Ward identity
one has to solve an integral equation for the full vertex.
 For the  spin vertex $\Gamma_s$ 
 we find the same result as above:
 the AL-type diagrams restore spin conservation.

Once we have established that AL blocks play the role of an effective single bosonic propagator, it is obvious
that these blocks also give additional
 contributions to the FL vertex $\Gamma^\Omega_{\alpha\beta;\gamma\delta} (\bk,0;\bp,0)$
and, hence, to Landau parameters. These contributions, shown Fig.~\ref{fig1} $c$), 
 add an extra term  of the structure
 $(3 \delta_{\alpha\delta} \delta_{\beta\gamma} - {\vec \sigma}_{\alpha\delta} \cdot {\vec \sigma}_{\beta\gamma}) F({\bf k}+{\bf p})$ to $\Gamma^\Omega_{\alpha\beta;\gamma\delta} 
 (\bk,0;\bp,0)$. Although 
 $ F({\bf k+\bp})$ is a rather complex function of its argument,
 one can show
that   
 its partial harmonics   are  the same as those of the RPA-like kernel
 $1/(\delta + a^2 ({\bf k}+ {\bf p})^2)$. 
This statement is true for harmonics with 
 $n \ll \lambda$ which are the only ones we need. Combining Eq.~(\ref{2_1}) with the AL contribution, 
  we obtain for the full vertex 
 near a FM QCP
  \bwt 
  \beq
\Gamma^{\Omega, \mathrm{full}}_{\alpha\beta;\gamma\delta} 
 (\bk,0;\bp,0)
 =  \frac{1}{2\nu Z}
 \left(\frac{{\vec \sigma}_{\alpha\delta} \cdot {\vec \sigma}_{\beta\gamma}}{\delta + a^2 ({\bf k}-{\bf p})^2} + \frac{3 \delta_{\alpha\delta} \delta_{\beta\gamma} - {\vec \sigma}_{\alpha\delta} \cdot {\vec \sigma}_{\beta\gamma}}{\delta + a^2 ({\bf k}+ {\bf p})^2} \right).
\label{3}
\eeq
\ewt

Two comments are now
  in order. First, the AL term in Eq.~(\ref{3}) 
 has both spin and charge components and the latter is by no means negligible. 
 In other words, {\it  a spin-conserving low-energy model contains singular interactions in both spin and charge channels}. 
 Second, the AL term has the same structure as the direct spin-fluctuation exchange term, but becomes singular at 
 small total momentum ($\bk\approx-\bp$)
 rather than at 
  small transferred momentum ($\bk\approx \bp$). 

 {\it Fermi liquid near a FM QCP.}~~~~
We now construct the new Landau function $g^{\mathrm{full}}_{\alpha \beta;\gamma \delta} (\bk,\bp) = Z^2 (m^*/m)
 \Gamma^{\omega,\mathrm{full}}_{\alpha \beta;\gamma \delta} (\bk,0;\bp,0) =
g^{\mathrm{full}}_c \delta_{\alpha \gamma} \delta_{\beta \delta} + g^{\mathrm{full}}_s 
{\vec \sigma}_{\alpha \gamma}\cdot {\vec \sigma}_{\beta \delta}$
 from the effective interaction $\Gamma^{\Omega,\mathrm{full}}$. After some standard spin algebra, we obtain
\bea
&&g^{\mathrm{full}}_{\alpha \beta;\gamma \delta} (\theta) = \frac{3}{2} \frac{\delta_{\alpha\gamma} \delta_{\beta\delta}}{\delta + 4 (a k_F)^2 \sin^2{\theta/2}} + 
\frac{1}{2} {\vec \sigma}_{\alpha\gamma}\cdot {\vec \sigma}_{\beta\delta} \times 
\nonumber \\
&& \left(
\frac{4}{\delta + 4 (a k_F)^2 \cos^2{\theta/2}} - \frac{1}{\delta + 4 (ak_F)^2 \sin^2{\theta/2}}\right),
\label{4}
\eea
where $\theta=\angle(\bk,\bp)$.
Evaluating $g^{\mathrm{full}}_{c,1}$ and $g^{\mathrm{full}}_{s,0}$, we obtain 
$g^{\mathrm{full}}_{c,1} = g_{s,0}^{\mathrm{full}} 
\approx \lambda$ for $\lambda\gg 1$, hence
\beq
m^*/m = \lambda; ~\chi_s  = \chi^0_s \frac{1 + \lambda}{1 +\lambda} = \chi^0_s.
\label{n_22}
\eeq
These two results are now in full agreement with 
  those of the loop-wise expansion, Eq. (\ref{n_1}).
\begin{figure}[tbp]
\caption{
(Color on-line.)  Schematic evolution of the partial components of the full 
Landau function [Eq.~(\ref{4})]  vs the inverse mass renormalization parameter $\lambda^{-1}$. All charge and even spin Landau components
are positive but diverge near a FM QCP, while odd spin components are negative. 
 The spin nematic component $g_{s,1}$ reaches the Pomeranchuk critical value of $-1$ first.}
\label{fig2}
\begin{center}
\epsfxsize=0.7\columnwidth
\epsffile{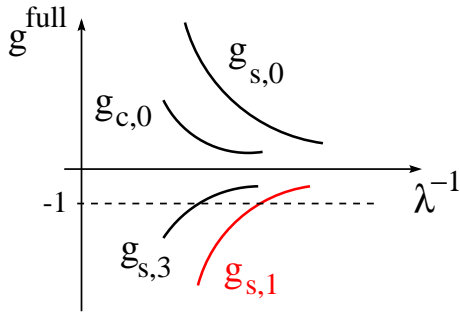}
\end{center}
\end{figure}

 The FL theory with the Landau function 
 from Eq.~(\ref{4}) is rather non-trivial. First, not only $g_{c,1}$ 
 but {\it all} charge components with $n \ll \lambda$ diverge near criticality in the same way: $g^{\mathrm{full}}_{c,n} = (3/2\pi)
 \int^\pi_0 d \theta \cos ({n
\theta})/(\delta + 4 a^2 k^2_F \sin^2{\theta/2}) \approx \lambda$ 
(see Fig. \ref{fig2}).
  The divergence of an infinite set of Landau parameters 
  also exists  near a charge QCP. \cite{we} 
This does not lead to dramatic consequences as the divergence of $g_{c,n>1}$ is compensated by that of $m^*/m=1+g_{c,1}$, and the charge susceptibilities $\chi_{c,n} = \chi^0_{c,n} (m^*/m)/ (1 + g^{\mathrm{full}}_{c,n}) =
\chi^0_{c,n} (1 + \lambda)/(1 + \lambda) = \chi^0_{c,n}$ remain intact. 

 Second, and more important, we see from Eq. ~(\ref{4})
 that all odd spin Landau parameters are negative 
 and diverge at the FM QCP. Since a Pomeranchuk instability occurs when the corresponding Landau parameter approaches $-1$, it implies that {\it a spin Pomeranchuk instability with finite angular momentum occurs  before the isotropic, Stoner-like instability}. Indeed, to leading order in $1/\lambda$, we have from 
 Eq.~(\ref{4}) 
 $g^{\mathrm{full}}_{s,n} =  \lambda \delta_{n,2m} - \frac{5}{3} \lambda \delta_{n,2m+1}$. For $n=2m+1$, $g^{\mathrm{full}}_{s,n} 
= -5 \lambda/3$, and $g^{\mathrm{full}}_{s,n}$ approaches $-1$ at
$\lambda = 3/5$. To discriminate between states with different angular momenta, one needs to evaluate the Landau parameters beyond the
  leading order
in $1/\lambda$. Doing so, we find, quite naturally, that $g^{\mathrm{full}}_{s,1}$  
approaches $-1$ first. Therefore, the leading instability corresponds to a $p$-wave spin nematic state.  Such a state in itinerant Fermi systems has recently attracted significant attention.~\cite{wu}
 Two different  phases
  ($\alpha$ and $\beta$) --the analogs of $A$ and $B$ phases of $3$He--
 have been identified, 
 but  only 
 the $\alpha$ phase is actually stable.
 This phase  is characterized by a $p-$wave spin order parameter 
$\Delta_s= \sum_{\bk} f(k)
(c^\dagger_{\bk\uparrow} c_{\bk\uparrow} - c^\dagger_{\bk\downarrow}
c_{\bk\downarrow}) \cos {\theta_{\bk}}$, where $f(k)$ is sharply peaked at the FS.
 This state has zero magnetization but the centers of FSs of spin-up and spin-down electrons are separated by finite momentum. 

As we have already mentioned, a $p-$wave 
 instability competes with more
 ``conventional'' instabilities of a FL near the FM QCP -- $p-$wave superconductivity, a first order transition and
 a transition into a helical magnetic phase. Which one 
 occurs first depends on the parameters of the model.  Controllable calculations are possible only
 when the range of interaction in the spin channel is large, i.e., for $ak_F\gg 1$. According to Ref.~\onlinecite{chi_last}, the
 most relevant out of ``conventional'' instabilities is the
 first order transition, which occurs at 
$\delta=\delta_{cr}^{1} \approx 0.21/(ak_F)$.
The $p$-wave Pomeranchuk 
instability occurs at   $\delta=\delta^{p}_{cr}= 25/16 (ak_F)^2$. Although $\delta^{p}_{cr}$ is smaller than  
 $\delta_{cr}^{1}$
 by power-counting, the 
 two critical values coincide at 
  large enough value of 
 $ak_F \approx 7.44$,  which
   implies that for a realistic case of
 $ak_F  \sim 1$, the $p$-wave instability 
 comes first.

Finally, we notice that $SU(2)$ spin invariance is 
  only approximate in real systems and 
broken by, e.g., spin-orbit (SO) coupling. This breaks the exact relation between the ladder
 and AL-type vertex corrections, but 
  the
 AL term in $\Gamma^\Omega$ 
 still remains of the same order as the bare spin interaction, and, as long as SO coupling is 
 small, the system still undergoes a $p-$wave spin-nematic instability at $\lambda  \sim 1$.   

To conclude, we considered the behavior of an 
$SU(2)$ invariant
itinerant fermionic system near a ferromagnetic quantum critical point. 
We showed that the spin-fermion model, traditionally used to describe  this behavior, does not satisfy $SU(2)$ spin conservation. We obtained a spin- and charge-conserving low-energy model for a FM QCP. This model has singular interactions in both spin and charge channels, as opposed to the SFM which considers interaction only in the spin channel. We constructed a FL theory for such a model and found that the system 
 undergoes a Pomeranchuk instability into a $p-$wave spin nematic state before a FM QCP is reached.

We acknowledge helpful discussions with 
C. Castellani, A. Green, I. Eremin, E. Fradkin, 
P. Hirsch, H.-Y. Kee, Eun-Ah Kim,  Y.-B. Kim, S. Kivelson, M. Lawler, A. Rosch, A. Varlamov, T. Vojta, and C. Wu, support from NSF-DMR-0604406 (A. V. Ch.) and NSF-DMR-0908029 (D. L. M.).

\end{document}